# Title

'Expertise and Task Pressure in fNIRS-based brain Connectomes'

# Authors:

F. Deligianni,[1*†], H. Singh, [1†] H.N. Modi,[1] S. Jahani,[2] M. Yücel,[2,3] A. Darzi,[1] D.R. Leff,[1†] G.Z. Yang[1,4‡]

# Affiliations:

[1]Hamlyn Centre for Robotic Surgery, Imperial College London, United Kingdom

[2] Neurophotonics Center, Biomedical Engineering, Boston University, Boston, MA 02215, USA

[3] Athinoula A. Martinos Center for Biomedical Imaging, Massachusetts General Hospital, Boston, USA

[4] Institute of Medical Robotics, Shanghai Jiao Tong University, China

[†]Joint First authors

# Corresponding authors:

[*]fani.deligianni@imperial.ac.uk, [†] d.leff@imperial.ac.uk, [‡] g.z.yang@imperial.ac.uk

# Acknowledgments

**Funding:** HS, HM, DRL and AD acknowledge support from the NIHR Imperial Biomedical Research Centre (BRC), FD acknowledges support from EPSRC (EP/R026092/1).

**Author contributions:** The study design was conceived and developed by HNM, HS, DRL, GZY and AD. HNM and HS executed the experiment and collected functional imaging data. Data pre-processing was conducted by HS, SJ, MY and connectivity analysis was conducted by FD. Data interpretation was performed by FD, HS and DRL in consultation with HNM, GZY and AD. The manuscript was drafted by FD, HS, HNM and DRL. Critical editing of the manuscript was performed by MY, SJ, GZY, AD, and DRL.

# Abstract

Acquisition of bimanual motor skills, critical in several applications ranging from robotic teleoperations to surgery, is associated with a protracted learning curve. Brain connectivity based on functional Near Infrared Spectroscopy (fNIRS) data has shown promising results in distinguishing experts from novice surgeons. However, it is less well understood how expertise-related disparity in brain connectivity is modulated by dynamic temporal demands experienced during a surgical task. In this study, we use fNIRS to examine the interplay between frontal and motor brain regions in a cohort of surgical residents of varying expertise performing a laparoscopic surgical task under temporal demand. The results demonstrate that prefrontal-motor connectivity in senior residents is more resilient to time pressure. Furthermore, certain global characteristics of brain connectomes, such as the small-world index, may be used to detect the presence of an underlying stressor.

# Keywords

# Introduction

Sensorimotor learning has attracted considerable interest among the neuroimaging community due to its importance in tasks that require bimanual dexterity and high levels of precision, such as surgical procedures, sports and teleoperations in extreme environments (1-5). Several studies have shown that cortical activation varies between groups of surgeons of different levels of experience (1, 3, 6). In particular, task evoked change in hemoglobin concentration has been found to attenuate across the prefrontal cortex (PFC) in senior surgeons compared to junior counter-parts (1, 6, 7). Moreover, progressive skills acquisition through seniority is associated with relative increase in hemoglobin concentration across sensorimotor regions such as supplementary motor area (SMA) and primary motor cortical areas (M1) (1, 8).

Optimally functioning executive centers in the brain are critical for technical success during surgical tasks (3). For example, selective attention – the ability of the brain to prioritize sensory information – plays an important role in managing competing mental demands. To some extent, training improves the ability to maintain motor dexterity and cognitive skills (e.g. decision-making) under time-pressure (9). However, regardless of experience, workflow disruptions increase surgeons' mental demands (10) which could prompt performance deterioration and threaten patient safety (9). In fact, it has been shown that there is an increase in performance with moderate workload, followed by deterioration following further increases in pressure. Subject-level tolerance to pressure and anxiety varies significantly across individuals (3, 11). Detecting workload and understanding how it modulates brain connectivity between prefrontal and sensorimotor regions could elucidate subject-specific susceptibility to errors and prevent performance failures during operative procedures.

Executive function and attention are supported by interconnected neural circuits that involve several brain regions which are significantly affected under stress and workload (12). In particular, the prefrontal cortex (PFC) is implicated in attention control, concentration, executive function and decision-making (3, 13, 14). Furthermore, ventromedial PFC and dorsomedial PFC are important for social processing and anxiety (15). Thus far, little is known about the functional connectivity between prefrontal and sensorimotor areas (16) and how this is modulated by high workload conditions experienced during complex surgical tasks.

Successful technical performance in surgery requires the integration of smooth motor control and cognitive skills such as attention and decision-making. Extensive research shows that technical skills assessment based on measures of performance such as task progression, leak volume, and errors do not relate well to operators' expertise (1, 8, 17). Furthermore, these measures cannot be translated from simulated to real scenarios, which is critical in order to measure the surgeon's resilience to stress online. Nevertheless, classification of surgeon expertise based on brain activation during simulated surgical tasks has been proposed as a robust way of distinguishing novices from experts (1, 8).

In this paper, the cortical hemodynamic changes that accompany motor and cognitive multitasking demands during a bimanual surgical task are investigated at brain connectivity level using functional near- infrared spectroscopy (fNIRS), Fig. 1. fNIRS has been employed to acquire relative changes in cortical haemodynamic data from both prefrontal and sensorimotor regions, during bimanual laparoscopic suturing ('key hole' stitching). Importantly, the paper presents changes in brain connectivity associated with expertise-level differences in surgeons' ability to cope with multiple competing demands. To our knowledge this is one of the first approaches to systematically study the dynamics of brain behaviour in surgeons operating in a naturalistic environment under a realistic temporal demand.

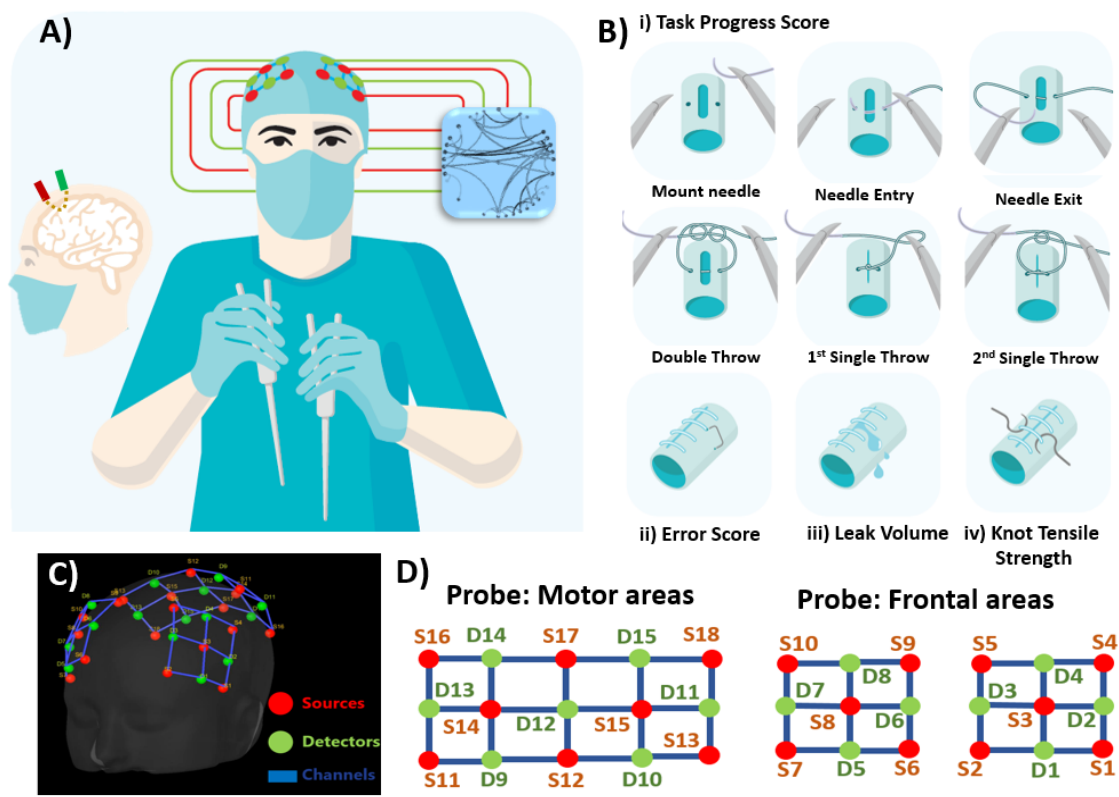

**Fig. 1: Surgical Training Task Design and Performance Assessment**: A) Schematic illustration of a surgical resident participant performing a laparoscopic suturing task according to the Fundamentals of Laparoscopic Surgery (FLS), B) Technical performance evaluation based on task progression, error score, leak volume and knot tensile strength, C) Relative positions of fNIRS optodes with respect to the underlying prefrontal and sensorimotor cortex, D) Probe configuration of the ETG-4000 fNIRS

array (Hitachi Medical Corp, Japan) of source emitters (red closed circles), detectors (green closed circles) and corresponding channels (blue lines).

## Materials and Methods

### Subjects

Twenty-nine surgical residents enrolled in the study (median age [range]=33 [29-57] years, 9 females) and were grouped based on their level of training into 'Junior', 'Intermediate', and 'Senior' cohorts, Table I. All participants were screened for handedness, and neuropsychiatric illness (n=0), and were asked to abstain from alcohol and caffeine intake for 24 hours prior to participation. Local ethical approval has been obtained (LREC:05/Q0403/142).

| TABLE 1. Demographics | | | | |
|---|---|---|---|---|
| **Resident Subgroup** | **Grade** | **Number** | **Median Age (Range)** | **Median Handedness Score (Range)*** |
| Junior | PGY1-2 | 11 | 31 (29 – 35) | 0.90 (-0.10 – 1.00) |
| Intermediate | PGY3-4 | 10 | 32 (30 – 37) | 0.70 (-0.10 – 1.00) |
| Senior | PGY5+ | 8 | 43 (38 – 57) | 0.70 (0.30 – 1.00) |

LS: laparoscopic suturing; PGY: postgraduate year. Previous LS experience refers to the number of times the task has been performed in the past. *Based on the Edinburgh Handedness Inventory

### Laparoscopic Suturing Task

Participants were instructed to perform a laparoscopic suturing (LS) task using an intracorporeal technique on a laparoscopic box trainer (iSim2, iSurgicals, UK). The task involved inserting a 2-0 Vicryl® suture (Ethicon, Somerville, NJ) as close to pre-marked entry and exit points on either side of a defect in a Penrose drain. To tie a knot laparoscopically, participants were instructed to formulate one double throw followed by two single throws of the suture.

**Experimental design**

All subjects performed the task under four experimental conditions: (1) 'self-paced', in which residents were permitted to take as long as required to tie each knot (Motor Self-Paced, MSP), (2) 'time pressure', in which, a two-minute per knot time restriction was applied, (Motor Time Pressure, MTP), (3) 'self-paced' knot tying whilst simultaneously answering a medical decision-making scenario (Motor-Cognitive Self-Paced, MCSP), and (4) 'time pressure' knot tying with a decision-making scenario (Motor-Cognitive Time-Pressure, MCTP).

A block design experiment was conducted in which participants performed the laparoscopic suturing task five times under each experimental condition with 30 second inter-trial rest periods between each knot. The order in which the subjects experienced the conditions was randomised.

The purpose of the decision-making scenarios was to impose a realistic cognitive demand on the subjects during the laparoscopic suturing task. The scenarios were vignettes of acute surgical patients who would be typically encountered during an "on-call" shift. The vignettes were pre-recorded and played at the same stage during the suturing task for each subject (when the needle was mounted on the needle holder). The participants were required to listen to the scenario whilst they continued suturing and give a brief response as to how they would manage the patient. Two expert surgeons rated the difficulty and authenticity of each scenario to ensure internal consistency and face validity of all the vignettes. Rating was based on the combined evaluation of the 'difficulty' and the 'authenticity' of each scenario on a scale of 1-10. Inter-rater reliability was tested using correlation (difficulty: $\rho=0.891$, $p<0.01$; authenticity: $\rho=0.835$, $p<0.01$).

**Functional Brain Imaging**

Functional near infrared spectroscopy (fNIRS) is a non-invasive functional neuroimaging technique which measures cortical absorption of near infrared (NIR) light to estimate the local concentration changes of oxygenated haemoglobin ($HbO_2$) and deoxygenated haemoglobin (HHb). The typical haemodynamic response of brain activation comprises a task-evoked increase in $HbO_2$ and a lower amplitude decrease in HHb. An ETG-4000 Optical Topography System (Hitachi Medical Co, Japan) was used to measure activation across 24 prefrontal cortical locations ('channels') and 22 motor cortical locations. The positions of which were defined according to the international 10-20 system of probe placement (18), as illustrated in Fig. 1.

**Pre-processing of Functional Neuroimaging Data**

Optical data was pre-processed as previously described (3) using a customized MATLAB-based toolbox (HOMER2) (19). Prefrontal and motor raw data streams were combined using bespoke MATLAB functions. After application of a low-pass filter (0.5Hz) to minimise high frequency noise and electrocardiographic effects on the data, 45 out of a total of 696 channels were excluded from further analysis due to poor optical signals, giving a data rejection rate of 6.5%. The remaining channel data was de-trended to correct for baseline fluctuations and averaged across blocks to increase the signal-to-noise ratio. Raw mean intensity values were converted to changes in optical density relative to the mean of each channel across the whole task period. Channel-wise motion detection and spline correction were performed using hybrid spline correction and Savitzky–Golay filtering (20). Relative changes in light intensities were converted into changes in $HbO_2$ and $HHb$ concentration using the modified Beer-Lambert Law. The resting time period (30sec) from each trial was removed and remaining time series were segmented into non-overlapping segments.

**Connectivity Estimation of Brain Networks Derived from NIRS**

Typically, large scale functional connectivity derived from neuroimaging data is modeled as a network or a graph. The nodes of the graph represent brain regions and, in this case, each region or "node" is associated with an fNIRS measurement channel.

We characterized brain connectivity from both $HbO_2$ and $HHb$ concentration data based on partial correlation, which is a measure of synchrony that disentangles the influence of indirect connections on the connectivity between each pair of brain regions (21, 22). In other words, it provides a measure of the signal transmitted directly between two regions. Since, it is estimated based on the inverse covariance matrix, it is important to form a well-conditioned, positive definite matrix. Finally, the geodesic mean of the covariance matrices of HbO$_2$ and HHb represents the overall connectivity matrix for each segment.

The covariance matrix of the fNIRS signal is estimated based on the 'Oracle Approximating Shrinkage' (OAS) estimator. In high dimensional problems of small sample number, shrinkage methods, such as the Ledoit Wolf (LW) and OAS help to better condition the covariance matrix and in general result in improved performance. OAS has demonstrated better performance than LW in functional neuroimaging data, especially when the number of samples is less than the number of dimensions (23).

**Comparison of Brain Networks based on NBS**

Network Based Statistics (NBS) was employed to identify "edges" that exhibit significant statistical differences between brain networks (24). This method alleviates the problem of false discovery rate, often observed in univariate edge-wise hypothesis tests. NBS utilizes clustering to identify subgraphs and associates the hypothesis test with a global network measure. In particular, a t-test is estimated for each graph edge and thus a thresholded graph is constructed. Finally, connected edges are identified as components and a p-value is computed by comparing with the null distribution. In this way, NBS provides greater power to detect differences between networks.

NBS is used to examine the strength of brain connectivity in senior and junior surgical residents during MSP, MTP, MCSP and MCTP conditions. We also use NBS to examine whether there are significantly stronger or weaker connections between MSP and MTP conditions among juniors, intermediate and senior surgical residents.

The analysis concludes with measures of global network efficiency and transitivity to derive the small-world index for weighted undirected graphs. The global efficiency, $E_g$ is related to the clustering coefficient and it is the average inverse shortest path length in the network (25). Transitivity, $T_g$ is the ratio of triangles to triplets in the network. A small world index is defined based on the ratio: $SWI = (E_g * T_g)/(E_r * T_r)$. Where $E_r$ and $T_r$ reflects the global efficiency and transitivity of random networks of equivalent degree distribution (26).

**Comparison of Brain Networks based on Predictive Models of Connectivity**

Predictive models of brain networks have been used in multi-modal comparisons of connectivity to identify both similarities and differences between sets of multi-modal brain networks (22, 27). This approach exploits sparse canonical correlation analysis to identify linear relationships between two sets of brain graphs (21, 28). Functional brain connectomes are projected onto a tangent space of the covariance manifold, and thus the predicted brain networks are constrained to symmetric positive definite matrices, which improves prediction performance. Sparsity parameters are optimized to maximise the ability to predict one dataset from the other and they are exploited to set a threshold statistic and thus enable connections to be identified that have a statistically significant relationship between the two datasets/conditions.

We exploit this methodology to examine the relationship between brain networks of the same subjects across two conditions, namely time pressure and self-paced. Sparsitiy parameters are estimated based on cross-validation, which subsamples brain networks 100 times, while randomized lasso is employed to permute the lasso parameters for 1000 times. Therefore, we perform 100000 random permutations for each comparison. Subsequently, we use the binomial distribution to define a threshold statistic based on the average value of the sparsity parameters. The distribution provides with how likely each connection is to be selected by chance or rejected by chance.

**Multivariate Classification based on Global Connectome Properties**

Global connectivity features SWI, efficiency and transitivity was used in a 10-fold cross-validation classification framework. Classification is based on an error-correcting output codes (ECOC) approach (29). ECOC classification combines binary Support Vector Machine (SVM) classifiers to form a four-classes classification across conditions and three-classes classification across expertise, respectively. SVM classifier has been built based on a Radial Basis Function (RBF) kernel. Fig. 5a-b shows 2D plots with relation to the efficiency and SWI of the color-coded conditions (MSP, MTP,

MCSP & MCTP) and expertise (juniors, intermediate, and seniors residents), respectively. Across conditions there is a profound separation of the classes that it is also reflected in the accuracy of the classifier that is 95.7% across classes based on a 10-fold cross-validation. Separation of expertise based on global connectome properties achieves accuracy of 76.06% based on a 10-fold cross-validation. The corresponding confusion matrices are plotted at Fig. 5c and Fig. 5d, respectively.

## Results

**Surgical Training Task Design and Performance Assessment**

To identify changes in brain connectivity that occur with intraoperative stress and whether these changes can be used to reliably detect when a surgical resident is under pressure, a cohort of juniors intermediate and senior surgical residents were recruited to perform a simulated laparoscopic suturing task while cortical hemodynamic data (an indirect measure of brain function) were captured using fNIRS. Simulated laparoscopic suturing is typically employed in the Fundamentals of Laparoscopic Surgery (FLS) program to assess surgical residents' performance. Surgical residents are classified into juniors when they are $1^{st}$ to $2^{nd}$ year surgical residents (PGY1-2), intermediate if they are within $3^{rd}$ to $4^{th}$ year surgical residents (PGY3-4) and seniors when they have more than five years of surgical training (PGY5+). This is the first study to report technical performance scores based on task progression, leak volume, error score and knot tensile strength simultaneously acquired with neuroimaging data that encompass both prefrontal and sensorimotor cortical regions. Fig. 1 shows a schematic illustration of the surgeon performing an FLS task along with the fNIRS probe configuration. Furthermore, it demonstrates the steps of the task along with objective measures of technical performance.

The FLS benchmark of surgical assessment is not designed to emulate real-surgical scenarios and thus it is difficult to identify whether a given resident can cope effectively under pressure or not. In order to understand the impact of time-pressure on technical performance, we employed a block-design experimental paradigm to acquire data during time pressure and self-paced conditions. In the Motor-Self-Paced (MSP) condition participants were required to tie a knot without any external pressure, whereas in Motor-Time-Pressure (MTP) participants are required to complete the knot-tying task within two minutes. There is a rest period of 30 sec between each task. To further intensify the cognitive demand, two more conditions were designed that required the participant to simultaneously tie a knot while responding to a medical decision-making scenario under no time constraint, Motor-Cognitive-Self-Paced (MCSP) and under a two minutes time constraint, Motor-Cognitive-Time-Pressure (MCTP).

The effectiveness of the design in imposing workload conditions is measured based on both subjective and objective scores (Appendix: Technical Skill Assessment). Subjective workload measures are based on the Surgery Task Load Index (SURG-TLX), whereas objective measures are based on task progression score, error score, leak volume and knot tensile strength. Task progression score is shown in Fig. 1 A-i and according to this measure surgical residents are rated from zero to six according to how far they have progressed in well-defined tasks: (1) mounting the needle onto the needle holder, (2) needle insertion into the drain, (3) exiting the needle from the drain, (4) double throw, (5) $1^{st}$ single throw, and (6) $2^{nd}$ single throw of a laparoscopic reef knot.

As illustrated in Fig 1 A-ii, error score (mm) is defined as the summed distances between the needle insertion or exit points and the pre-marked target positions on entry and exit, respectively. Leak volume (ml) illustrated in Fig. 1 A-iii, is introduced to simulate a bleeding vessel and reflects the volume of saline leaking from the closed surgical knot repair site over one minute. Finally, the knot tensile strength, highlighted in Fig. 1 A-iv, quantifies the tensile strength of each knot, which is important in surgical procedures to evaluate how knot security.

Subjective workload measures, such as Surgery Task Load Index (SURG-TLX) demonstrate significant differences between self-paced and time pressure conditions only in junior surgical residents (Appendix: Fig. A1). However, some objective performance scores reveal statistically significant differences between self-paced and time-pressure tasks that are apparent in intermediate and senior surgical residents (Appendix: Fig. A2, Fig. A4). Regarding progression scores, statistically significant differences were observed between self-paced and time-pressure conditions in juniors (p-val<0.001) and in intermediate surgical residents (p-val<0.01). Regardless of expertise group, statistically significant differences in leak volume are observed between conditions based on within-subjects statistical analysis (p-val<0.001). On the other hand, error scores and tensile strength failed to show within-subject differences for any of the groups of junior, intermediate and senior surgical residents. Please, note that within-subjects differences reflect statistical analysis across conditions within each group of expertise with the same subjects, participating in all conditions.

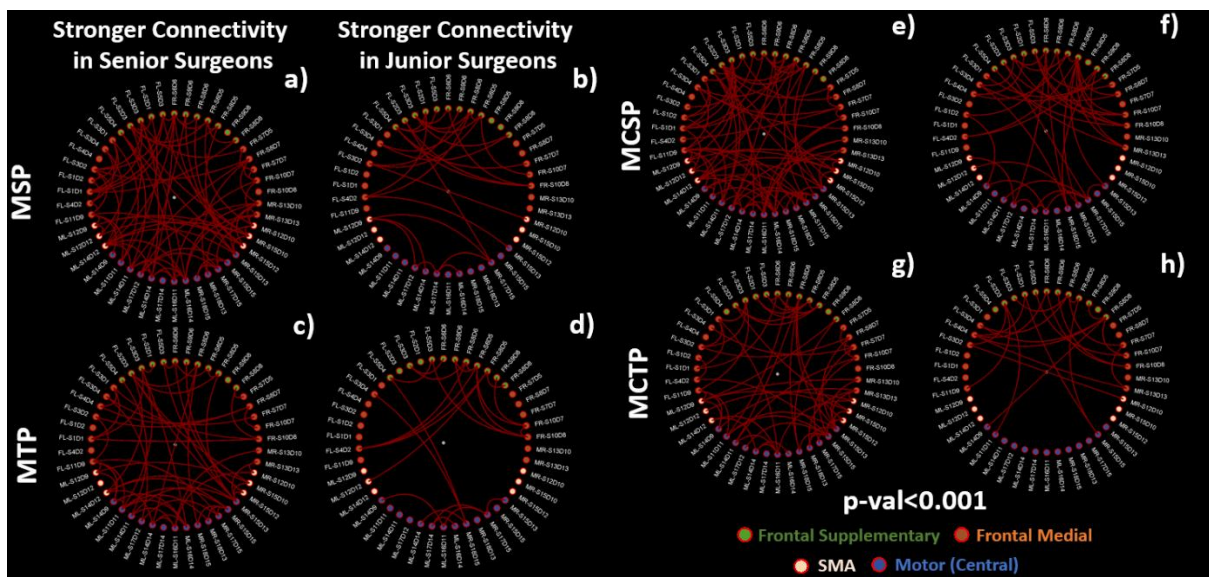

**Fig. 2: Expertise-related differences in brain connectivity:** Connectome results of Network based statistics applied in Motor Self-Paced (MSP), Motor Time-Pressure (MTP), Motor-Cognitive-Self-Paced (MCSP) and Motor-Cognitive-Time-Pressure (MCTP) conditions. For each condition two hypothesis testing based on t-test explored: stronger connectivity in senior residents compared to the other groups and stronger connectivity in junior residents compared to the other groups. Statistically significant between-group differences in connectivity strength are represented with red lines for each hypothesis testing. Each node in the connectivity maps, represents a NIRS channel. These channels are sorted based on their location from interior to posterior and from left to right. Therefore, they cover Frontal-Left (FL), Frontal-Right (FR), Motor-Left (ML) and Motor-Right (MR) brain regions, respectively. The color-coding of the nodes, green, orange, white and blue represent Frontal

Supplementary (green), Frontal Medial (orange), SMA (white) and Motor Central (blue) brain regions, respectively.

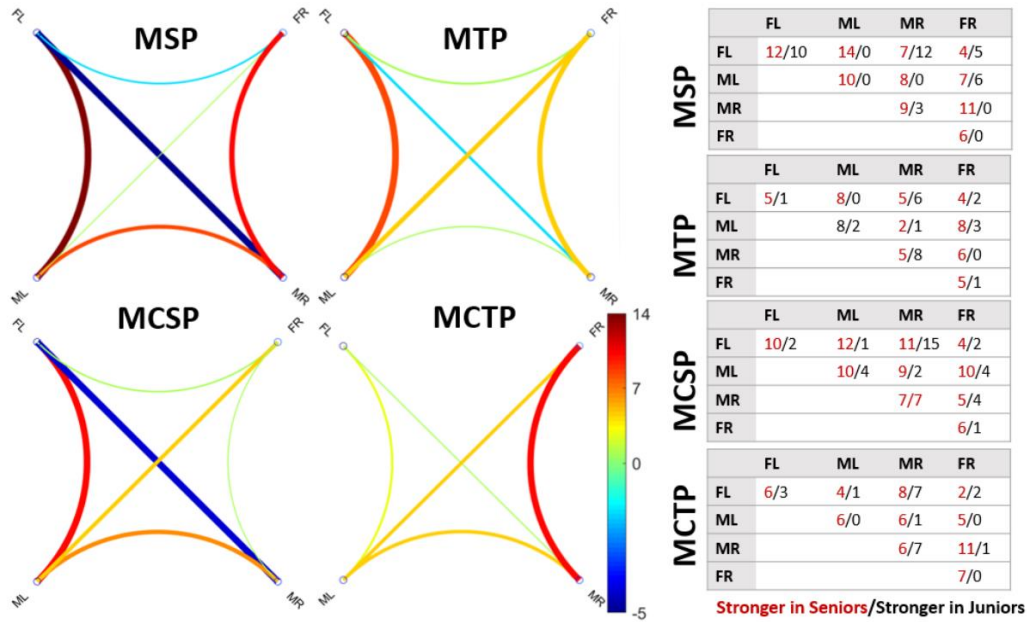

**Fig 3: Differences across conditions between junior and senior surgical residents in the number of brain connections that link Frontal and Motor Regions.** Graphical illustration depicting differences in the number of stronger connections in senior and junior residents that link Frontal Left (FL), Frontal Right (FR), Motor Left (ML) and Motor Right (MR) regions. Strong intra-hemispheric connectivity is observed between Frontal and Motor regions across all conditions. Junior residents demonstrate stronger inter-hemispheric connectivity between FL and MR regions.

**Expertise and Task-pressure Related Changes in Brain Connectivity**

We identified statistically significant expertise-related differences in brain connectivity in each of the four experimental conditions, MSP, MTP, MCSP and MCTP, as highlighted in Fig. 2. Statistically significant differences in network connectivity has been identified in examining the hypotheses: i) whether connectivity is stronger in seniors compared to the other groups, ii) whether connectivity is stronger in juniors surgical residents compared to the other groups. To better understand how these differences vary across expertise and condition we cluster channels into Frontal Left (FL), Frontal Right (FR), Motor Left (ML) and Motor Right (MR). Senior residents demonstrate stronger intra-hemispheric connections than junior residents in all experimental conditions, as illustrated in Fig. 3. On the other hand, junior residents display stronger inter-hemispheric connections between left frontal and right motor brain regions. The quantitative differences of brain connectivity between seniors and junior surgical residents became smaller in time-pressure condition compared to self-paced condition.

We also used network-based statistics to examine within-group differences across workload conditions and only amongst junior surgical residents exist statistically significant differences between the MSP and MTP conditions observed (p-val<0.01), as illustrated in Fig. 4 a). Nevertheless, significant covariations emerge between MSP vs MTP and MCSP vs MCTP conditions in juniors and seniors surgeons, respectively, Fig. 4b). These covariations have been estimated based on predictive models of connectivity. Each subfigure in Fig. 4b) shows the intersection of connections that are common for both predicted and predictive variables. We observe that the highlighted connections are much higher in number in senior surgeons than in juniors.

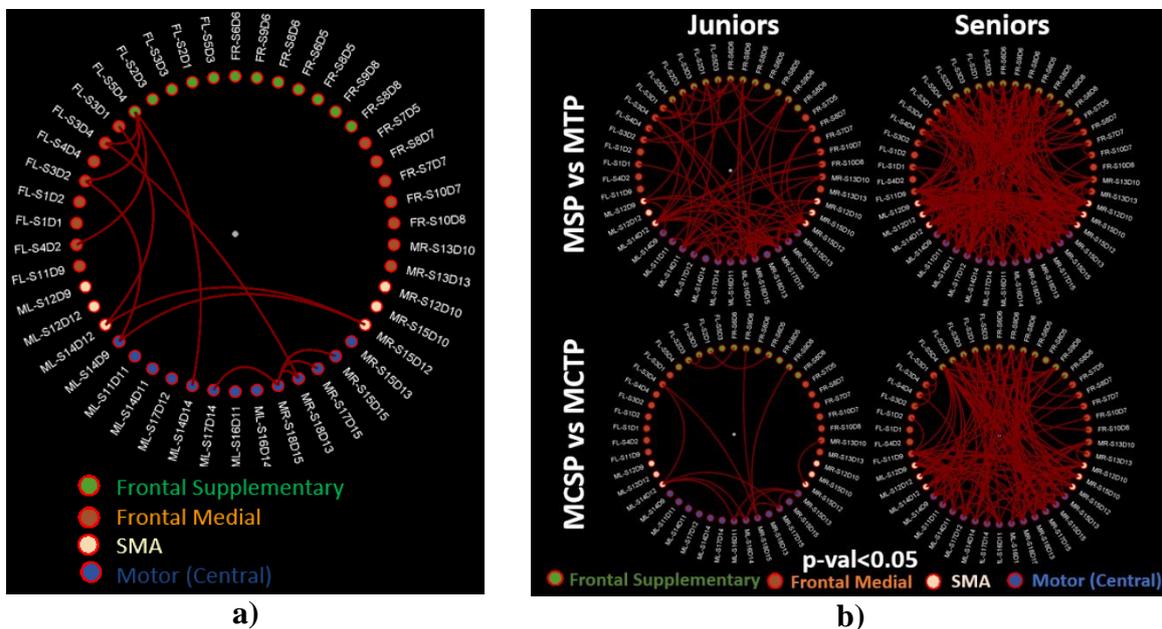

**Fig. 4. Time Pressure-Related Changes in Brain Connectivity**. A) Based on NBS, statistically significant differences in brain connections between time-pressure and self-paced conditions were identified only for junior surgeons between MTP vs MSP. B) Statistically significant covariations between MSP and MTP networks and MCSP and MCTP networks were identified both in juniors and senior surgeons based on predictive models of brain connectivity. Fig. 4 b) shows connections that covary significantly across MSP vs MTP and MCSP vs MCTP conditions in juniors and seniors surgeons, respectively. We observe that these covariations are much higher in number in senior surgeons than in juniors. This confirms that brain networks under bimanual tasks in senior residents are more resilient to time pressure compared to junior surgeons.

**Global Connectome Properties Predict both Expertise and Time-Pressure**

'Small-world' properties in brain connectomes reveal network topological differences that affect the efficiency of information integration between distant brain regions. Therefore, the Small-World-Index (SWI) is a measure of network efficiency and it has been shown to be able to encode differences between functional brain networks derived from neurological or psychiatric disease and normal controls (25). In this study the SWI increases significantly from juniors to intermediate to seniors across all conditions. Fig. 5 demonstrates how the SWI, global efficiency and transitivity compare across different levels of surgical expertise (juniors, intermediate, seniors) in MSP, MTP, MCSP and MCTP conditions. The SWI in senior residents is statistically greater than in both juniors and intermediate residents (p-val<0.001). Critically, time-pressure affects SWI, in both the motor task as well as the combined motor-cognitive tasks and the differences are profound.

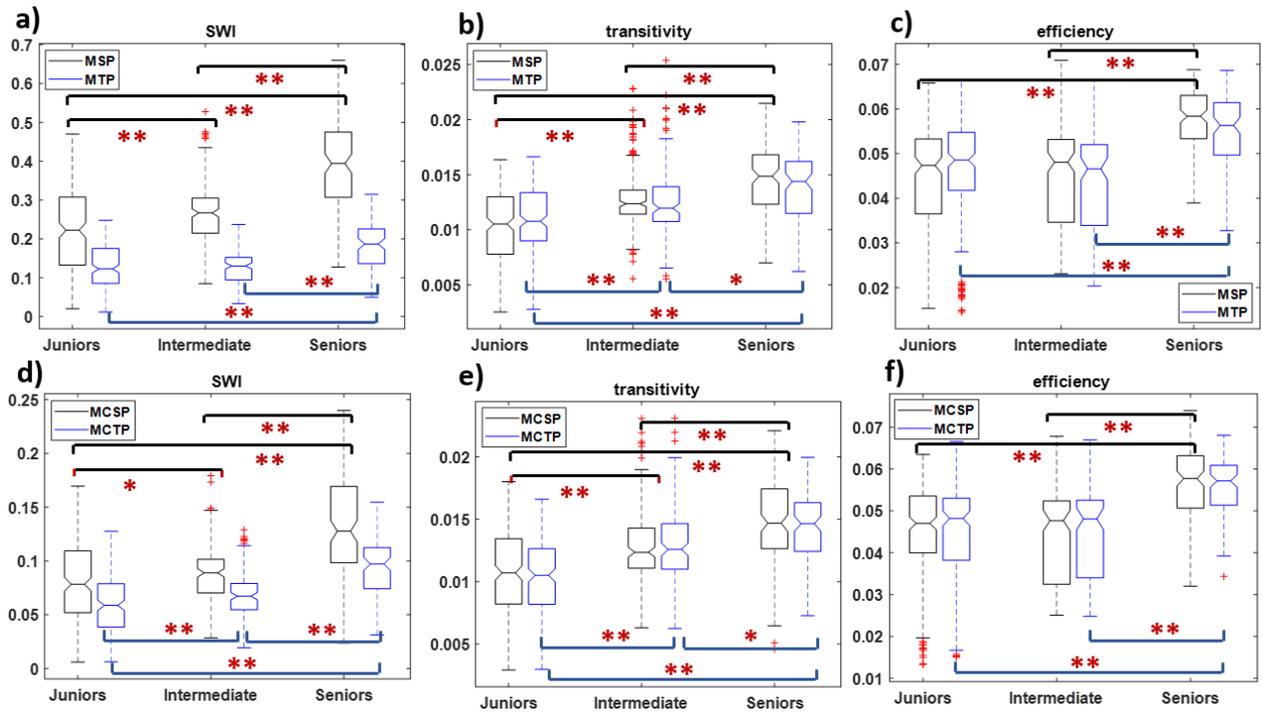

**Fig. 5. Global properties of brain networks across conditions and expertise.** Box and whisker plots depicting Small-World Index (SWI), global efficiency and transitivity. Comparisons in network econometrics across the different levels of surgical training (juniors, intermediate, seniors) in MSP, MTP, MCSP and MCTP conditions are highlighted. The red asterisks denote statistical significance with * for p-values<0.05 and ** for p-values<0.001.

Key challenges in bimanual skills assessment are the ability to evaluate and distinguish surgical residents that are better able to cope under pressure with relation to their training. SWI, transitivity and efficiency are global properties of brain connectomes and in this study they have been used to classify and predict the underline experimental condition. The results show that the accuracy across experimental condition exceeds 95.7% based on a 10-fold cross validation, whereas classification accuracy across expertise is 76.06%. The confusion matrices are depicted in Fig. 6 and they show that the largest error in classification across conditions occurs between MCTP and MCSP conditions, whereas the largest error across expertise occurs between juniors and seniors surgical residents.

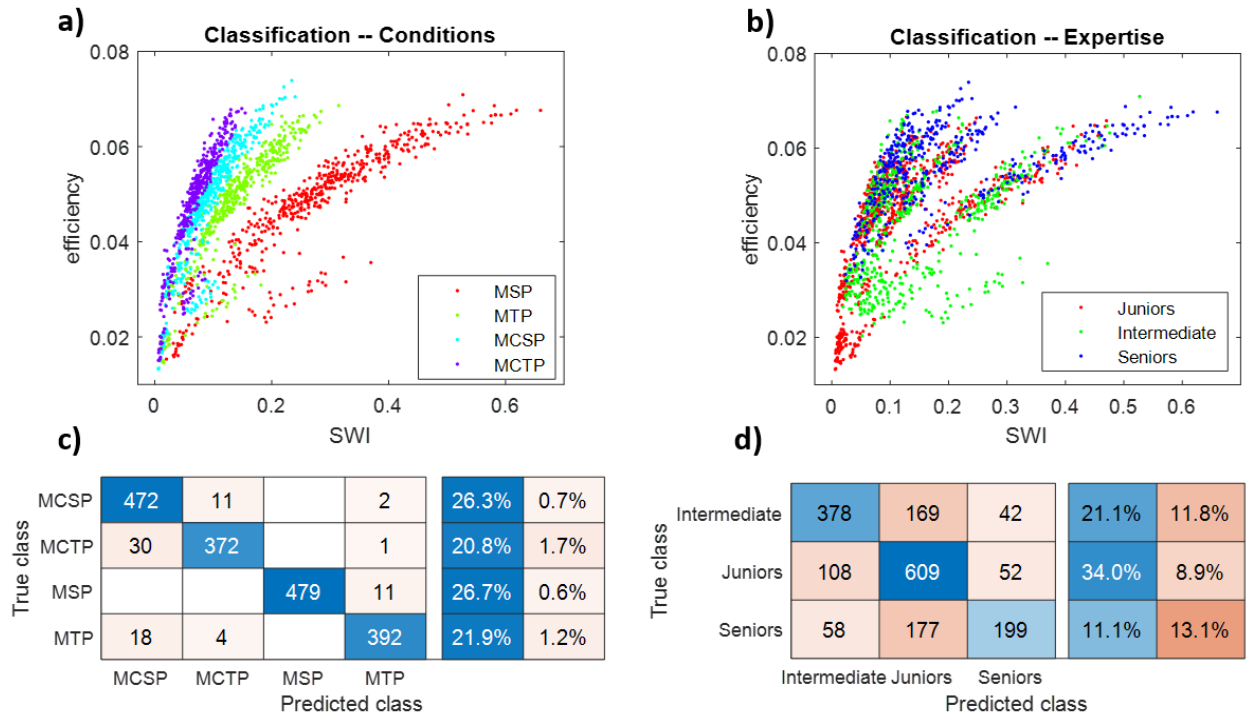

**Fig. 6. Classification results across experimental condition and across expertise**. An error-correcting output codes (ECOC) classifier for multi-class classification based on SVM (Radial basis function kernel) has been employed. A) Classification features of small-world-index (SWI) and efficiency color-coded according to the experimental condition reveal profound separation across conditions, B) SWI and efficiency features color-coded according to expertise show how senior residents score higher in terms of both SWI and efficiency. C) Performance of the classification across experimental conditions based on the confusion matrix. The overall accuracy of the classifier based on a 10-fold cross-validation is 95.7%, D) Performance of the classification across expertise based on the confusion matrix. The overall accuracy of the classifier is 76.06% based on 10-fold cross validation.

# Discussion

Understanding the way in which bimanual skills acquisition changes brain connectivity is the subject of extensive research in neuroscience, neurorehabilitation, and human-machine interaction (30-32). Surgeons face a plethora of stressors in the operating room which can be detrimental to technical performance and potentially jeopardize patient safety (33). Thus, it is crucial for surgeons to maintain motor dexterity as well as decision-making ability during stressful operating conditions (3). Recent studies attest to relative attenuation in prefrontal function amongst surgeons performing under temporal stress (3). However, other non-primary cortical motor regions, such as the SMA and premotor cortex are also involved in motor planning and execution (34). In fact, there is evidence of strong interaction between the motor system and working memory, which is the process of retaining information in mind (34, 35). Working memory modulates spatial attention (12) and it is significantly affected under acute psychological stress (36). Moreover, non-primary motor networks may support working memory under increased cognitive demands (34, 37). The current study is important because it is the first to investigate the interaction between PFC and sensorimotor regions in surgeons under more realistic temporal pressure.

Furthermore, brain activation during motor training is also modulated by task complexity and learning time (3). Therefore, the neural response to intraoperative stress should be studied in relation to operator's experience. Laparoscopic suturing is particularly difficult to master because it requires extensive hand-eye coordination and visuospatial skills, and execution is hampered by loss of depth perception and the fulcrum effect. Learning of surgical skills can take weeks to months of training, whereas perfection of the skill requires several years of experience in the operating room. Prefrontal cortical activation diminishes with practice of simple surgical tasks (e.g. open knot tying) within few sessions of training; whereas in more complex tasks, (e.g. laparoscopic suturing), attenuation in prefrontal activity takes several weeks of practice (8). Fast motor skill learning, in the course of single training sessions has been associated with interactions between Supplementary Motor Area (SMA), preSMA and sensorimotor areas (31). Later phases of learning that occur over the course of several weeks/months have been associated with activation at M1, SMA and the cerebellum (38).

Using a block design experiment, this fNIRS study delineates the changes in brain connectivity that occur under temporal demand in a cohort of surgical residents with varying experience performing a laparoscopic suturing and/or decision-making task. The stark differences in brain connectivity between junior and senior surgical residents reflect the neural adaptation to stress that occurs with years of training. This is demonstrated both with network-based statistics that aim to identify differences per connection that are statistically significant as well as global connectome properties. The small world index reveals that functional organization in senior residents is more efficient during both self-paced and time-pressure conditions. However, the small world index becomes significantly smaller in time pressure in both seniors and juniors indicating less efficient network topology under task pressure, regardless of expertise. SWI has been very popular in neuroimaging and neuroscience because it has a neurobiological interpretation and it provides a global measure of network topology (25). High values of SWI reflect regional specialization with efficient information transfer between remote channels/regions (26).

The variation in brain connectivity across expertise is consistent with literature that highlights significant differences in prefrontal cortical activation as a function of surgical expertise (1, 3, 6). Whilst traditional performance metrics are often used to distinguish surgical expertise, studies attest that the relationship between surgeon's experience and technical ability as measured using these metrics is not clear (39). Exposing differences in brain function between expert and novice surgical residents may provide a more robust method of benchmarking skill acquisition. Critically, brain connectivity measures can distinguish novice from expert surgical residents with a high degree of accuracy (1, 40).

A delicate balance between functional intra-hemispheric and inter-hemispheric connectivity drives motor network performance during bimanual tasks. In a recent study of the effects of ageing on the motor network, it has been shown that phase synchronization between motor left and motor right regions is anticorrelated with bimanual performance (41). Here, we demonstrate stronger intra-hemispheric connectivity between frontal and motor regions in senior residents, and stronger inter-hemispheric connectivity between left frontal and right motor regions in junior residents. Differences in the number of connections between senior and junior residents diminish under time pressure in both motor and motor-cognitive conditions.

Network-based analysis identified statistically different connections only between MTP and MSP in juniors, which is consistent with the hypothesis that senior surgical residents are more resilient to stress and thus differences would be more difficult to detect. Nevertheless, global connectome properties have demonstrated robust statistically significant differences across conditions that indicate significant differences in the network topology. It is challenging to compare networks with varying number of nodes and density (42). To alleviate this problem, we compare the topology of weighted networks/graphs with edges that are statistically different across expertise and the same number of nodes across networks.

In summary, bimanual surgical tasks require extensive training to enhance dexterity and acquire smooth motor control. Over the course of several months or years, surgeons learn to operate efficiently and become more resilient to environmental stressors. Performance depends on workload and mental demands as well as operator proficiency. fNIRS monitors hemodynamic brain changes that relate to brain function and has been shown to be particularly successful in evaluating the proficiency of surgeons. Furthermore, stress and performance degradation have been associated with attenuation of brain activity in prefrontal cortex. To our knowledge, the interplay between motor and frontal regions is not well understood because it requires more complex fNIRS probe configuration. This is one of the first studies to show that brain connectivity in senior residents is more 'efficient' than in juniors, and that time-pressure results in a significant decline in network efficiency in both seniors and junior residents.

# Appendix

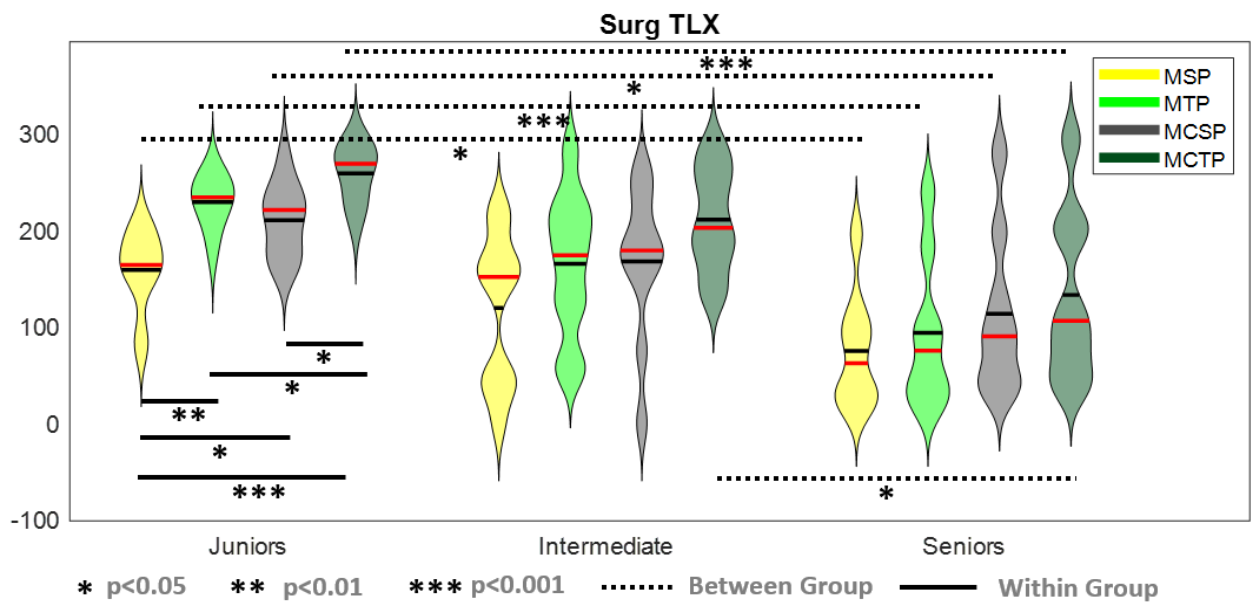

**Fig. A1: Subjective workload measured based on SURG-TLX.** Violin plots for each condition (MSP, MTP, MCSP, MCTP) and each expertise level (Juniors (PGY1-2), Intermediate (PGY3-4), Seniors (PGY5+)). Statistically significant within-group and between-group differences are highlighted.

**Technical Skill Assessment**

Workload has been subjectively quantified based on SURG-TLX data, whereas technical skills were objectively assessed using four performance parameters similar to previous work (3): Task Progression Score (TPS; arbitrary units, au), Error Score (mm), Leak Volume (ml), and Knot Tensile Strength (KTS; Newtons, N).

Statistical analysis was performed using SPSS version 23.0. Within-group comparisons were analysed using ANOVA with repeated measures along with post hoc tests using the Bonferroni correction for parametric data, such as SURG-TLX and leak volume. The Friedman test with post hoc Wilcoxon Signed Rank test was used for non-parametric data, such as task progression score, error score and knot tensile strength. For each condition, one-way ANOVA with Tukey post hoc tests (parametric data) or the Kruskal-Wallis and post hoc Mann-Whitney U tests (non-parametric data) was used to determine significant between-group differences (i.e. PGY1-2 vs. PGY3-4 vs. PGY5+) in workload or technical performance. The Shapiro-Wilk test has been used to test normality in each dataset of performance measure.

Fig. A1 demonstrates statistically significant within-group and between-group differences in subjective workload measured based on SURG-TLX. Within-class comparisons reveal statistical significant differences only within Junior surgical residents. Between-group comparisons demonstrate statistically significant differences between Seniors and Junior surgical residents in each corresponding pair of experimental condition. Significant differences between Senior and Intermediate surgical residents exist only for the MCTP condition.

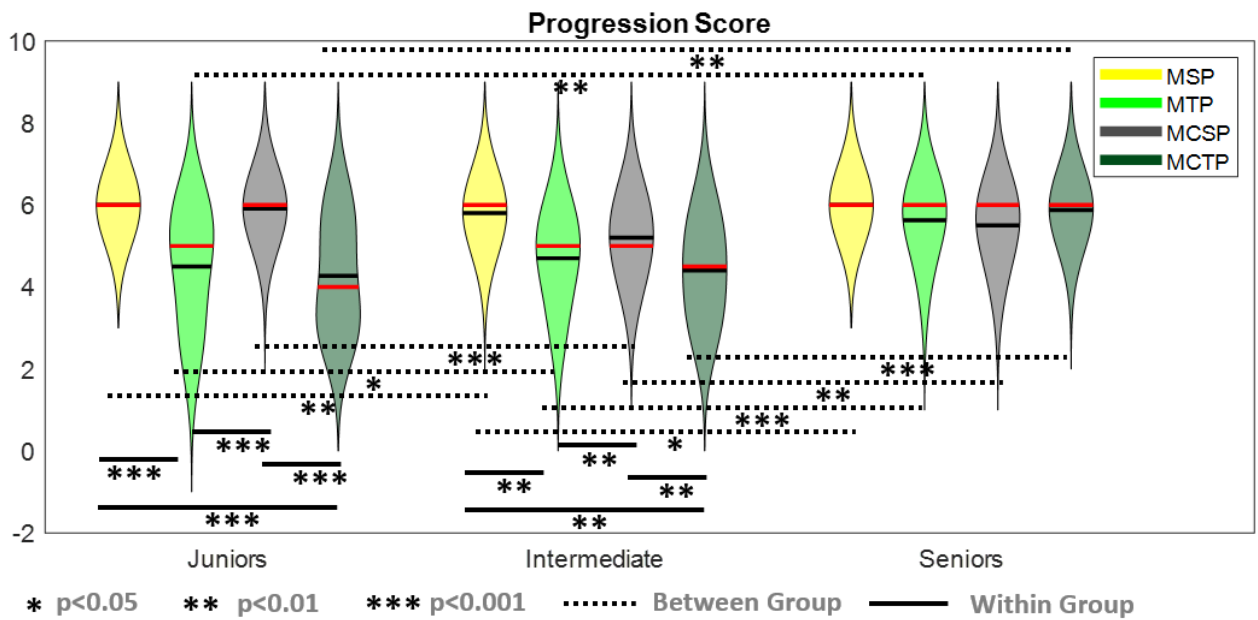

**Fig. A2: Technical skill assessment based on progression score.** Violin plots for each condition (MSP, MTP, MCSP, MCTP) and each expertise level (Juniors (PGY1-2), Intermediate (PGY3-4), Seniors (PGY5+)). Statistically significant within-group and between-group differences are highlighted.

Fig. A2 demonstrates statistically significant within-group and between-group differences in objective skill assessment based on progression score. Within-class comparisons reveal statistically significant differences both for Junior and Intermediate surgical residents but not for Senior surgical residents. Between-group comparisons show statistically significant differences between Seniors and Intermediates for each corresponding pair of conditions. Significant differences between Junior and Senior surgical residents exist in MTP and MCTP conditions only.

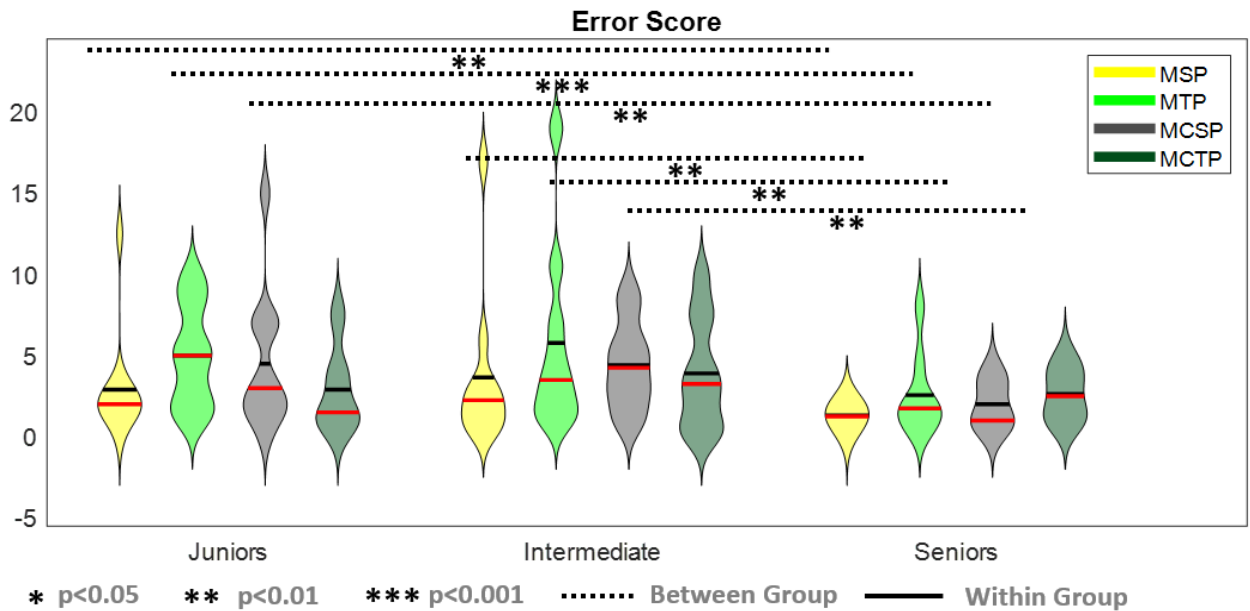

**Fig. A3: Technical skill assessment based on error score.** Violin plots for each condition (MSP, MTP, MCSP, MCTP) and each expertise level (Juniors (PGY1-2), Intermediate (PGY3-4), Seniors (PGY5+)). Statistically significant within-group and between-group differences are highlighted.

Fig. A3 demonstrates statistically significant between-group differences in objective skill assessment based on error score. Significant differences were identified between Senior and Junior surgical residents in MSP, MTP and MCSP condition. Significant differences were identified between Senior and Intermediate surgical residents in MSP, MTP and MCSP condition. No statistically significant within-group differences were identified.

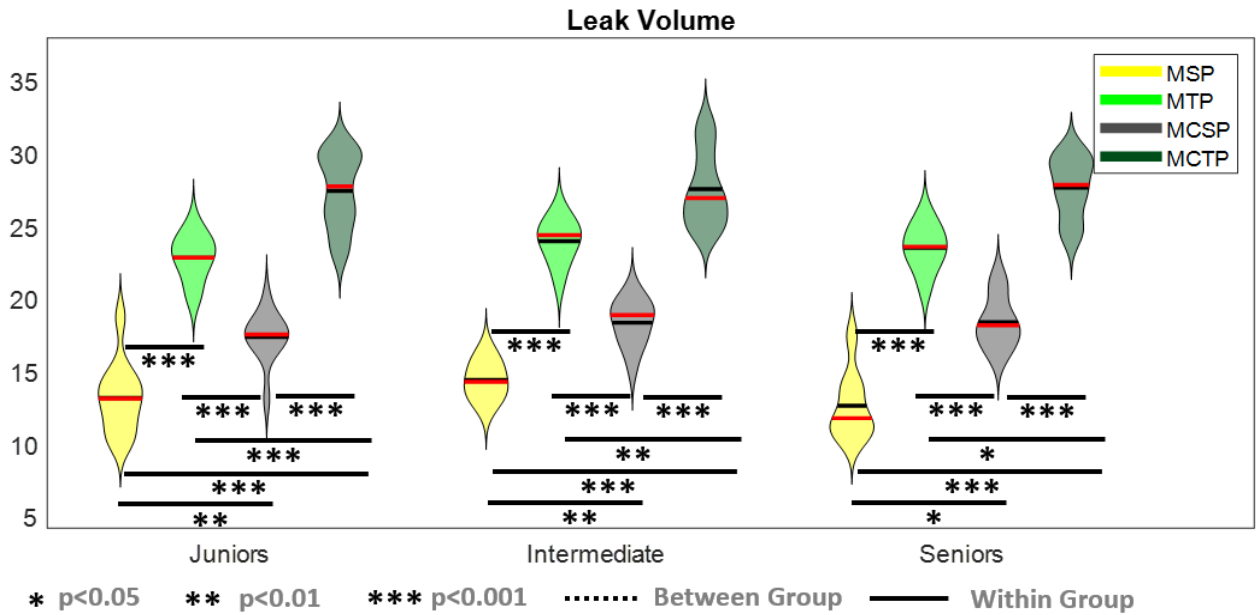

**Fig. A4: Technical skill assessment based on leak volume.** Violin plots for each condition (MSP, MTP, MCSP, MCTP) and each expertise level (Juniors (PGY1-2), Intermediate (PGY3-4), Seniors (PGY5+)). Statistically significant within-group and between-group differences are highlighted.

Fig. A4 demonstrates statistically significant within-group differences in objective skill assessment based on leak volume. Significant differences were identified between self-paced and time pressure conditions as well as between single task (motor) and dual task (motor and cognitive) across all levels of expertise. No statistically significant between-group differences were identified.

No statistically significant within-group or between-group differences were identified based on the knot tensile strength.